# Problem of optimization of a transport traffic at preliminary registration of queires with use of CBSMAP-model.

Kondrashova E.V.[1]

**Abstract.** The problem of optimization of a transport traffic at preliminary registration of demands with use of the CBSMAP model is investigated. For the solution of an objective application of the queueing theory and the theory of controlled processes is supposed.

## INTRODUCTION

Recently problems of optimization of transport traffic become more actual. Traffic problem: hours-long traffic jams, congestion of roads, increase in road accident – all this leads to difficulty of the traffic, not economy of personal and working hours of participants of the transport traffic. In big cities, such as Moscow, optimization due to construction or expansion of roads in certain sites of the city becomes technically impossible . Thus there is interesting a traffic optimization problem definition (cars, buses, trucks, etc.) at preservation of roads without additional construction and expansion, due to management of a transport traffic and a task of certain criteria for participants of the traffic.

Object of control in a control system of traffic is the transport flow/traffic consisting of technical means of the cars, public transport which are carrying out freight transportation of objects, etc.)

## 1. The major factors influencing on control of transport traffic

Notice that at control of transport traffic, first, it is necessary to be guided by some integral characteristics of the transport traffic, and secondly, it is necessary to set some criteria of optimization, being guided on which it is possible to formulate and solve a problem of optimum control.

At control of city transport traffic the following characteristics are usually considered: not-stationarity of transport streams, stochasticity, incomplete controllability, coherence of several factors at the same time (road accident number, a delay in way, the average speed of the movement, etc.)

The essence of incomplete controllability consists what even in the presence of full information on traffic and possibility of informing drivers on necessary actions, these requirements have advisory nature. Thus the problem of control becomes complicated. Existence of reliable information and forecasting of behavior of drivers and load of roads' zones at certain time doesn't help to facilitate a road situation and leads to traffic jams. This situation is especially characteristic for the Moscow roads.

[1] Department of Applied mathematics, Moscow State Institute of Electronics and Mathematics, National Research University Higher School of Economics, Moscow, Russia, elizavetakondr@gmail.com

## 2. Control and optimization of car traffic with use of Internet service of electronic registration

Now various Internet services becomes more popular, such as programs-navigators, maps, services for registration and submission of documents by citizens, signing up in policlinic on an appointment, tracking of turns at the moment in facilities of paperwork, various mobile services, etc. These services are especially convenient to that the participant of this service becomes known either recommendatory time, or strictly determined time of visit, movement regulated for it, etc.

We will give examples of the recommendatory and regulated time for the participant of service.

At implementation of signing up in policlinic with use the Internet service such as emias.info, pgu.mos.ru, the participant receives the coupon for a certain date and time. This system allows not to create large turns and assumes saving of time of the participant, and also other participants of the system. Thus time strictly regulated for this participant and depending on this time the time intervals regulated for other participants of service are provided.

In the presence of program-service-navigator various ways of movement are offered to the participant of traffic (the best or the worst that depends on movement time to the specified destination). In this case service provides the recommendation which can adhere, or to choose other route.

In a case when optimization of traffic depends on "stochastic" or predictable preferences of drivers control becomes rather complex challenge. To predictable preferences of movement can be referred: seasonal movement, movement within a day (morning, day, evening, night), dependence on a day of the week (week or the day off), etc.

Despite the fact that it is possible to call preferences predictable, the possession of the information not always allows to optimize the traffic and the participant of the traffic comes up against a situation of "load" of roads. It is possible to refer spontaneous, unplanned trips to "stochastic" preferences.

For optimization of certain roads' zones it is offered to apply the Internet services to creation of the regulated time interval for each participant. Creation of such time intervals, first, facilitates optimization of a transport traffic for zones of roads without carrying out additional construction and expansion of the carriageway (where it is impossible), secondly, does traffic model more predictable and provides to participants reliable information about load not at the moment of time, and in advance, on future time interval.

## 3. The scheme of queries with Internet-service use

Give an example of work of service for a zone of the road, for entrance on which participants submit applications/queries for a certain time. Let us say to the participant of traffic in a certain day (perhaps, in some limited time interval) it is necessary to use a site of Road A1 at which the system of electronic admissions coupons will work for movement. The participant uses Internet service and, as well as other participants, submits the application for entrance.

The site of the road gets out (perhaps, the highway). Further the participant date and estimated time of entrance to the zone A1 gets out. The system contains data on the applications/queries, which are already submitted earlier, and "models" a traffic situation. In case traffic density doesn't exceed some preset value and the traffic on a site/zone A1 won't be complicated, to the participant the electronic coupon (electronic permission) to number of the car driving in the zone A1 is issued.

If for required time at addition of additional participants traffic is complicated, the system will offer the coupon on entrance to the zone A1 on other time interval, the closest to the interval requested by the participant at which choice the traffic won't be complicated. Thus at a strict regulation of time the transport stream/traffic will be settled also the traffic on the set road zone, using control, will become optimum.

It is necessary to consider in addition some problems, which can arise at a choice of control at the expense of a regulation of Internet service.

In a case when the participant needs to get on a zone of Road A1, but the system doesn't offer electronic coupons (for example for the reason that a large number of participants are already registered on the next time intervals, and addition of new participants will result in load of the road), the participant of Internet service can or choose a way of a detour for this time (another route), or get the coupon on paid entrance to a zone.

In this regard, there is a problem of calculation of volume of spare "bunker" of paid demands for time intervals. That is issue of paid coupons should not influence deterioration of a road situation and load of the road due to acquisition of the paid coupon.

At present, the system of paid roads is already familiar to citizens on the example of the toll highway Moscow-St. Petersburg and creation of paid roads to the airports. Paid roads in foreign countries are also widely used.

It should be noted that the road situation in Japan was improved due to construction of a large number of additional high roads the traffic on which is paid. However, there are also free ways of a detour.

However introduction of the control system and optimization is guided generally not by introduction of a fare for certain roads' zone, and by regulation of intensity of the traffic for the account of granting reliable information to participants about future condition of the traffic intensity and traffic ability on the set zone.

Calculation of borders of a time interval is also necessary for each road zone on which service will be used. Because hit in a temporary point of an interval will be casual (as the motorist cannot get to the destination in strictly certain time at the expense of the factors influencing traffic), it is necessary to consider it in mathematical model. And for various roads the admissible interval can appear different length: 10 minutes, 30 minutes, etc.

**CONCLUSION**

For modeling an expansion of the CBSMAP-model developed earlier and application of management to this model will be used.

**Pic.1 Flowchart of registration of the participant for entrance to a zone**

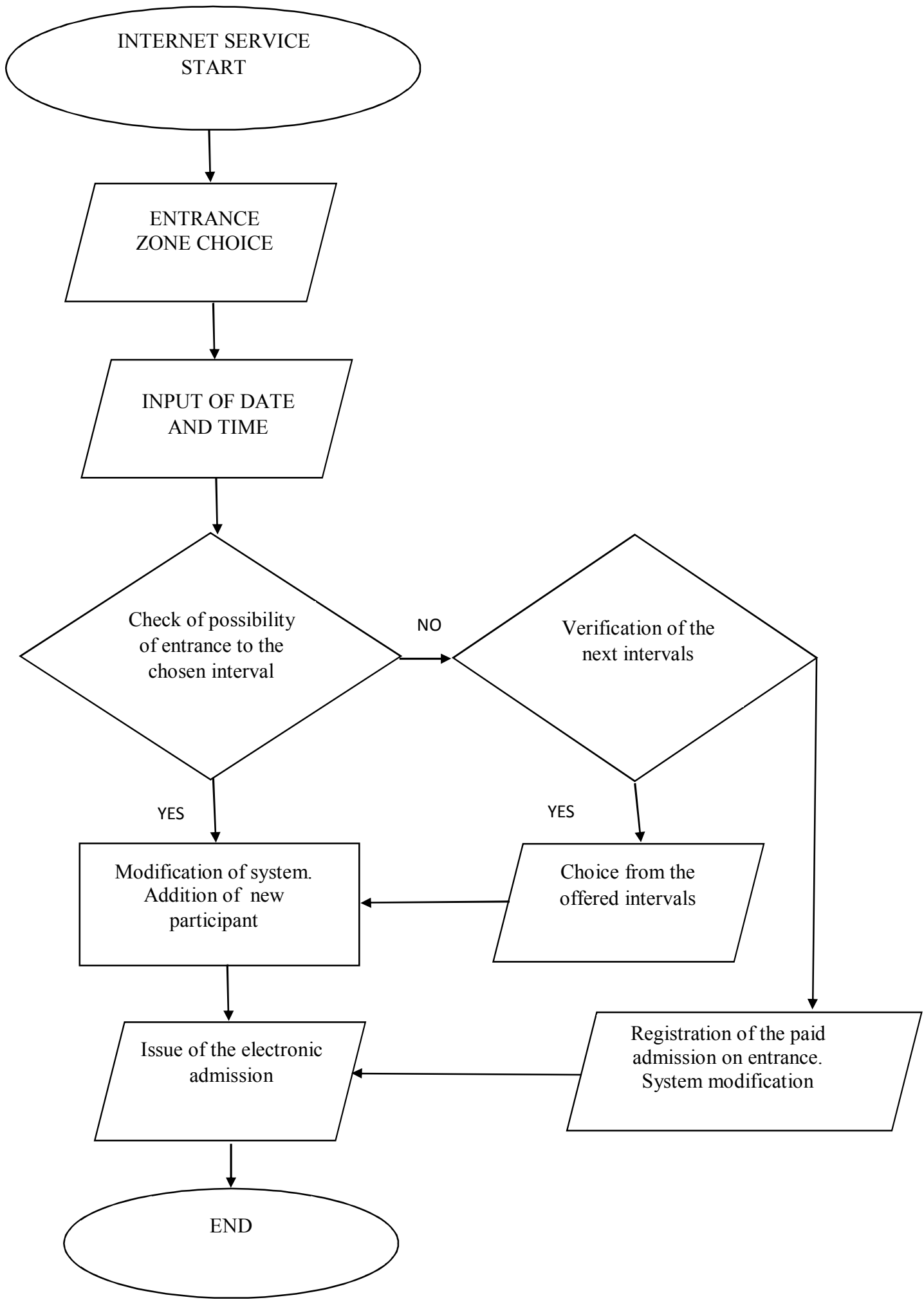